\documentclass[10pt, doublecolumn]{IEEEtran}
\usepackage{epsfig,latexsym}
\usepackage{float}
\usepackage{indentfirst}
\usepackage{amsmath}
\usepackage{amssymb}
\usepackage{times}
\usepackage{subfigure}
\usepackage{psfrag}
\usepackage{cite}
\usepackage{lastpage}
\usepackage{fancyhdr}
\usepackage{color}
 \usepackage{amsthm}
\usepackage{bigints}
\newtheorem{Proposition}{Proposition}
\newtheorem{Lemma}{Lemma}
\newtheorem{Corollary}{Corollary}
\newtheorem{lemma}[Lemma]{$\mathbf{Lemma}$}
\newtheorem{proposition}[Proposition]{Proposition}
\newtheorem{corollary}[Corollary]{$\mathbf{Corollary}$}
\begin{document}
\title{ {\huge  NOMA Meets Finite Resolution Analog Beamforming in Massive MIMO and Millimeter-Wave  Networks }}

\author{ Zhiguo Ding, \IEEEmembership{Senior Member, IEEE}, Linglong Dai, \IEEEmembership{Senior Member, IEEE},        Robert Schober, \IEEEmembership{Fellow, IEEE},

and
H. Vincent Poor, \IEEEmembership{Fellow, IEEE}\thanks{
Z. Ding and H. V. Poor   are with the Department of
Electrical Engineering, Princeton University, Princeton, NJ 08544,
USA.   Z. Ding is also with the School of
Computing and Communications, Lancaster
University, LA1 4WA, UK.  L. Dai is  with the Tsinghua National Laboratory for Information
Science and Technology, Department of Electronic Engineering, Tsinghua
University, Beijing, China. R. Schober is with   the Institute for Digital
Communications, University of Erlangen-Nurnberg, Germany.
}\vspace{-2em}}
 \maketitle
\begin{abstract}
Finite resolution analog beamforming (FRAB) has been recognized as  an effective approach   to reduce   hardware costs in massive multiple-input multiple-output (MIMO) and millimeter-wave networks. However, the use of FRAB means that the  beamformers are not perfectly  aligned with the   users' channels and multiple users may be assigned similar or even identifical beamformers. This letter shows how   non-orthogonal multiple access (NOMA) can be used to exploit this feature of FRAB, where a single FRAB based beamformer is shared by multiple users. Both analytical and simulation results are provided to demonstrate the excellent performance achieved by this new NOMA transmission scheme.
\end{abstract}\vspace{-2em}
\section{Introduction}
Non-orthogonal multiple access (NOMA) is a promising multiple access technique for   next generation wireless networks \cite{nomama,7434598x}, and  has been shown to be compatible  with many important 5G technologies, including  massive multiple-input multiple-output (MIMO) and millimeter wave (mmWave) transmission \cite{Zhiguo_massive, 7504208, Zhiguo_mmwave}.

A recent development in massive MIMO and mmWave networks is the use of finite resolution analog beamforming (FRAB), which reduces hardware costs \cite{7394147, Daiicc16}. Analog beamforming  does not alter the amplitude of a signal, but  modifies its phase only, which is different from digital beamforming. The finite resolution constraint on analog beamforming is due to the fact that the number of  phase shifts supported by a practical  circuit  is finite \cite{7394147, 7448838}.  An example for one-bit resolution analog beamforming is provided  in Table \ref{s}.  Depending on the values of   a user's complex-valued channel coefficients, $1$ or $-1$ will be chosen as the beamformer coefficients, as shown in Table I. The reduced hardware costs of FRAB  are at the expense of   performance losses  since the   obtained  beamformers are not perfectly  aligned with the target users' channels.

The purpose of this letter is to demonstrate that   the   characteristics  of FRAB   favour  the use   of  NOMA.
    Consider again  the example shown in Table \ref{s}. To clearly show the benefit of the combination of FRAB and NOMA, the users' channel vectors are chosen to be orthonormal, i.e., the users' channel vectors are normalized and orthogonal to each other.  The base station constructs  two beams according to the channel state information (CSI) of the two users in $\mathcal{S}_1$. If digital beamforming with perfect resolution was used, the beamforming vector for user $1$ in $\mathcal{S}_1$ would be  simply this user's channel vector, and therefore this beamformer could not be used by the two users in $\mathcal{S}_2$, since  this beamformer would be orthogonal to the two users' channel vectors. On the other hand, if FRAB is used, the formed two  beams are no longer   orthogonal to the two users'  channel vectors. In fact, for  the special case shown  in Table I,  the beamformer preferred by user $i$ in $\mathcal{S}_1$ is exactly the same as that of user $i$ in $\mathcal{S}_2$, even though the two users  have orthogonal channel vectors. As a result, NOMA has been applied to ensure that all the four   users can communicate concurrently. In this letter, a new NOMA transmission scheme that  exploits the features  of FRAB is proposed, and analytical results for the corresponding outage probabilities and diversity  gains of the users   are presented. The provided simulation results   demonstrate not only the excellent  performance of the proposed NOMA scheme, but also   the accuracy of the developed analytical results. We note  that the developed  analytical results concerning  the diversity gains are also applicable to conventional MIMO scenarios  without   NOMA, and hence shed light on  the performance loss caused by FRAB in a general MIMO network.

\begin{table}
  \centering
  \caption{An example for   finite resolution analog beamforming  }
  \begin{tabular}{|c|c|c|c|c|}
\hline
& user $1$ in $\mathcal{S}_1$ & user $2$ in $\mathcal{S}_1$&user $1$  in $\mathcal{S}_2$ &user $2$ in $\mathcal{S}_2$\\
    \hline
    channel& -0.19 + 0.66j	&-0.49 + 0.16j	&-0.27 - 0.11j &	-0.33 + 0.25j \\ vectors&
-0.06 - 0.53j	&-0.35 + 0.22j &	-0.06  + 0.58j &	-0.45 + 0.10j
\\ &0.34 - 0.03j	&-0.10 - 0.62j	&0.31 - 0.05j &	-0.20 + 0.59j
\\ &0.31  - 0.18j &	-0.06 + 0.41j &	0.34 - 0.60j	&-0.45 - 0.15j
\\
    \hline
   FRAB&  -1 &  -1 &  -1 &  -1\\beam-
  &-1 &   -1 &  -1 &   -1\\ formers &1 &  - 1 &  1 &   -1\\  &1 &  - 1 &  1 &  -1\\\hline
  \end{tabular}\vspace{1em}\label{s}\vspace{-3em}
\end{table}\vspace{-1em}
\section{System Model}
Consider a NOMA downlink scenario, in which  the base station is equipped with $M$ antennas. Assume that there are two groups of single-antenna users in the network. Denote by $\mathcal{S}_1$ a set of   users with strict quality of service (QoS) requirements, whose distances to the base station are denoted by $d_{yk}$ and are assumed to be fixed. Denote by $\mathcal{S}_2$ a set of users to be served opportunistically, and these users are uniformly distributed  in a disk-shaped area with radius $r_1$, where the base station is  at the center.  Denote the distances of the users in $\mathcal{S}_2$ to the base station by $d_{xi}$.  The $M\times 1$ channel vector of a user in $\mathcal{S}_1$ ($\mathcal{S}_2$) is denoted by $\mathbf{h}_k$ ($\mathbf{g}_i$). Two types of channel models are considered in this paper, namely  Rayleigh fading and  the mmWave model \cite{Daiicc16}, where the mmWave channel vector is modelled  as follows:
\begin{align}
\mathbf{h}_k =\frac{a_k}{1+d_{yk}^\alpha} \begin{bmatrix} 1 & e^{-j \pi \theta_k} &\cdots e^{-j \pi (M-1)\theta_k}\end{bmatrix}^T.
\end{align}
Here,  $\alpha$ denotes the path loss exponent, $\theta_k$ is the normalized direction, and $a_k$ denotes the fading attenuation coefficient. Note that for the purpose of illustration, only the line-of-sight path is considered for the mmWave model. \vspace{-1em}
\subsection{Implementation of Finite Resolution Analog Beamforming}
Suppose  that the  users in $\mathcal{S}_1$ are served via FRAB.  Denote by $\mathbf{f}_k$ the $M\times 1$ beamforming vector for user $k$, where  each element of $\mathbf{f}_k$ is drawn from the following vector:
\begin{align}
\bar{\mathbf{f}} = \begin{bmatrix} 1 &e^{j\frac{2\pi}{N_q}}&\cdots  &e^{j\frac{(N_q-1)2\pi}{N_q}} \end{bmatrix},
\end{align}
where $N_q$ denotes the number of   supported phase shifts.

The $i$-th element of $\bar{\mathbf{f}}$ is chosen as the $m$-th element of $\mathbf{f}_k$ based on the following criterion:
\begin{align}
i^*_{k,m} = \underset{i\in\{1, \cdots, N_q\}}{\arg \min}  \quad  \left|\bar{ {f}}_i-\frac{h_{k,m}}{|h_{k,m}|}\right|^2,
\end{align}
where $\bar{f}_i$ denotes the $i$-th element of $\bar{\mathbf{f}}$, and $h_{k,m}$ denotes the $m$-th element of user $k$'s channel vector.\vspace{-1em}
\subsection{Implementation of NOMA}
To reduce the system complexity, suppose  that only one user from $\mathcal{S}_2$  will be  paired with user $k$ from $\mathcal{S}_1$ and denote this user by user $i^*_k$.
The base station sends a superposition of   the messages of the two users on each beam.   User $k$ in $\mathcal{S}_1$ treats its partner's message as noise and decodes its own message with the following signal-to-interference-plus-noise ratio (SINR):
\begin{align}
\text{SINR}_k = \frac{ |\mathbf{h}_k^H\mathbf{f}_k|^2\alpha_{0,k}^2}{ |\mathbf{h}_k^H\mathbf{f}_k|^2\alpha_{1,k}^2+ \underset{l\in\mathcal{S}_1\setminus k}{\sum}|\mathbf{h}_k^H\mathbf{f}_l|^2 +\frac{M}{\rho}},
\end{align}
where the factor $M$  in the denominator is due to  the transmit   power normalization, and  the power allocation coefficients are denoted by  $\alpha_{n,k}$. Note that $\sum^{1}_{n=0}\alpha_{n,k}^2=1$ and $\alpha_{0,k}\geq \alpha_{1,k}$.

 By applying successive interference cancellation (SIC), user $i^*_k$ can decode its partner's  message  with the following SINR: $
 {\text{SINR}}_k^{k\rightarrow i^*_k} = \frac{ |\mathbf{g}_{i^*_k}^H\mathbf{f}_k|^2\alpha_{0,k}^2}{ |\mathbf{g}_{i^*_k}^H\mathbf{f}_k|^2\alpha_{1,k}^2+ \underset{l\in\mathcal{S}_1\setminus k}{\sum}|\mathbf{g}_{i^*_k}^H\mathbf{f}_l|^2 +\frac{M}{\rho}}$.
Let  $\epsilon_i=2^{R_i}-1$, $i\in\{0,1\}$, where $R_0$  and $R_1$ denote the targeted data rates for user $k$ and user $i^{*}_k$, respectively. If $ {\text{SINR}}_k^{k\rightarrow i^*_k}\geq \epsilon_{0}$, SIC can be carried out  successfully at user $i_k^*$ and the SINR for decoding its own message is given by
\begin{align}
 {\text{SINR}}_k^{ i^*_k} = \frac{ |\mathbf{g}_{i^*_k}^H\mathbf{f}_k|^2\alpha_{1,k}^2}{   \underset{l\in\mathcal{S}_1\setminus k}{\sum}|\mathbf{g}_{i^*_k}^H\mathbf{f}_l|^2 +\frac{M}{\rho}}.
\end{align}

We use the following user selection criterion:
\begin{align}\label{selection}
i^*_k = \arg \max \{ {\text{SINR}}_k^{k\rightarrow 1}, \cdots ,  {\text{SINR}}_k^{k\rightarrow |\mathcal{S}_2|}\}.
\end{align}
Note that this criterion   selects that user which maximizes the probability of successful  intra-NOMA interference cancellation, a key stage for SIC. Since the users in $\mathcal{S}_2$ are served opportunistically, we allow one user from $\mathcal{S}_2$ to be included in more than one pair. More sophisticated user scheduling algorithms can be designed to realize   fairness for the users in $\mathcal{S}_2$, which are not presented here due to space limitations.

\vspace{-0.5em}
\section{ Performance Analysis}
To the best knowledge of the authors, the impact of FRAB on the diversity gain has not been analyzed yet, not even  for scenarios  without NOMA. In order to obtain insight into  the performance of the proposed NOMA scheme,  in this section, we focus  on the special case with  $N_q=2$, $|\mathcal{S}_1|=1$, and   Rayleigh fading channels. Note that $N_q=2$ represents the case of  one-bit resolution  analog beamforming \cite{7448838}.
\vspace{-1em}
\subsection{Performance of  the User in $\mathcal{S}_1$}
 When there is a single beam, i.e., $|\mathcal{S}_1|=1$, the outage probability achieved by  the user in $\mathcal{S}_1$ is given by
\begin{align}\nonumber
\mathrm{P}^o_k= &\mathrm{P}\left(|\mathbf{h}_k^H\mathbf{f}_k|^2<\phi_0\right),
\end{align}
 where $\phi_i=\frac{\frac{M\epsilon_i}{\rho}}{\alpha_{i,k}^2 - \epsilon_i\sum^{1}_{n=i+1}\alpha_{n,k}^2}$, $i\in\{0, 1\}$.  Note that  $\alpha_{i,k}^2 >\epsilon_i\sum^{1}_{n=i+1}\alpha_{n,k}^2$ is assumed in this paper, since otherwise the outage probability is always one.  In order to find the cumulative distribution  function (CDF) of  $|\mathbf{h}_k^H\mathbf{f}_k|^2$, the following proposition is provided first.
\begin{proposition}\label{proposition}
Consider $M$  independent and identically distributed (i.i.d.)  random variables, denoted by $z_m$, each of which follows the   folded normal distribution, i.e., $z_m$ is the the absolute value of a Gaussian variable with   mean $0$ and variance $\frac{1}{2}$. The CDF of $z_{\sum}\triangleq \left|\sum^{M}_{m=1} z_m\right|^2 $    can be approximated as follows:
  \begin{align}
 F_{z_{\sum}}(z)
\approx& \frac{2^M}{\pi^{\frac{M}{2}}}\frac{z^\frac{M}{2}}{M!},
 \end{align}
 when $z\rightarrow 0$.
\end{proposition}

The following lemma provides an asymptotic approximation for the CDF of  the effective channel gains of the user in $\mathcal{S}_1$.
\begin{lemma}\label{lemma1}
For user $k$ in $\mathcal{S}_1$, the CDF of its effective channel gain on   beam $\mathbf{f}_k$ can be approximated as follows:
\begin{align}
F_{|\mathbf{h}_k^H\mathbf{f}_k|^2}(y)
\approx& \frac{2^{M}\left[y(1+d^\alpha_{yk}) \right]^{\frac{M+1}{2}}B\left(\frac{3}{2}, \frac{M}{2}\right)}{\pi^{\frac{M}{2}}(M-1)!M^{\frac{1}{2}}\Gamma\left(\frac{1}{2}\right)} ,
\end{align}
for $y\rightarrow 0$, where $B(\cdot)$ denotes the Beta function and $\Gamma(\cdot)$ denotes the Gamma function.
\end{lemma}

By using Lemma \ref{lemma1} and with some algebraic manipulations,  the following corollary   can be obtained.
\begin{corollary}\label{corollary 1}
For the proposed  NOMA system with Rayleigh fading, user $k$ achieves a diversity gain of  $\frac{M+1}{2}$.
\end{corollary}
{\it Remark 1:} With $M$ antennas at the base station, the full diversity gain for the considered scenario is $M$, but only a diversity gain of $\frac{M+1}{2}$ is achieved by the proposed scheme. This performance loss is mainly due to the use of FRAB.

{\it Remark 2:} It is important to point out that Corollary \ref{corollary 1} is general and   applicable to  conventional MIMO networks without  NOMA as well, since  the users in $\mathcal{S}_1$ do not perform SIC.

\vspace{-1em}

\subsection{Performance of the User in $\mathcal{S}_2$}
When there is a single beam, the outage probability achieved by  the user in $\mathcal{S}_2$ is given by
\begin{align}
\mathrm{P}^o_{i^*_k}= &1 - \mathrm{P}\left(|\mathbf{g}_{i^*_{k}}^H\mathbf{f}_k|^2>\max(\phi_0,\phi_1)\right).
\end{align}
Note for the case of $|\mathcal{S}_1|=1$,   the proposed  user selection criterion shown in \eqref{selection} simplifies to:
\begin{align}\label{selection1}
i^*_k = \arg \max \{ |\mathbf{g}_1^H\mathbf{f}_k|^2, \cdots , |\mathbf{g}_{|\mathcal{S}_2|}^H\mathbf{f}_k|^2\},
\end{align}
since $f(x)\triangleq \frac{ x\alpha_{0,k}^2}{ x\alpha_{1,k}^2+ \frac{M}{\rho}}$ is a monotonically  increasing function of $x$. Therefore,  the CDF of $|\mathbf{g}_{i^*_k}^H\mathbf{f}_k|^2$ can be obtained   as follows.

Denote a user randomly chosen from $\mathcal{S}_2$ by  user $\pi(i)$, and  the fading and path loss components of its composite  channel gain, $ \mathbf{g}_{\pi(i)} $,  can be decomposed as  $\mathbf{g}_{\pi(i)}=\frac{\bar{\mathbf{g}}_{\pi(i)}}{1+d^\alpha_{x\pi(i)}}$.  Therefore,  the effective fading  gain of this user, $|\bar{\mathbf{g}}_{\pi(i)}^H\mathbf{f}_k|^2$,  is exponentially distributed, since $\bar{\mathbf{g}}_{\pi(i)}$ and $\mathbf{f}_k$ are independent and a unitary  transformation of a Gaussian vector is still Gaussian distributed. It is important to point out that $|\bar{\mathbf{g}}_{\pi(i)}^H\mathbf{f}_k|^2$ is exponentially distributed with   parameter   $\frac{1}{M}$, instead of $1$ as in \cite{Nomading}. By using this observation and also following steps similar to those in  \cite{Nomading},  the composite channel gain     has the following approximate CDF:
\begin{align}\label{Px}
F_{\pi(i)}(y) = \sum^{N}_{n=1}w_n   (1-e^{-c_n\frac{y}{M}}),
\end{align}
where $N$ is a parameter for the  Gauss-Chebyshev approximation, $w_n =  \frac{\pi}{2N}\sqrt{1-\eta_n^2} (\eta_n+1)$, $\eta_n=\cos\left(\frac{2n-1}{2N}\pi\right)$, and $c_n=1+\left(\frac{r_1}{2}\eta_n +\frac{r_1}{2}\right)^\alpha$.

 After applying the simplified criterion in \eqref{selection1} and also assuming  that the users' channels are i.i.d., the outage probability of user $i^*_k$  to decode its message delivered on beam $k$ is $\left(F_{\pi(i)}(\max(\phi_0, \phi_1))\right)^{|\mathcal{S}_2|} $. After some algebraic manipulations, we  obtain the following corollary.
\begin{corollary}
For the proposed  NOMA system with Rayleigh  fading, the full diversity gain  of  $|\mathcal{S}_2|$ is achievable by the user from $\mathcal{S}_2$.
\end{corollary}

\begin{figure}[t]
\begin{center}
\includegraphics[width=0.35\textwidth]{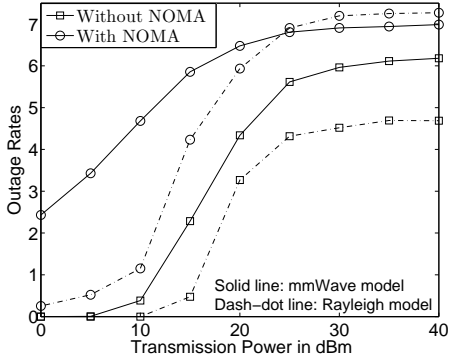}\vspace{-1em}
\end{center}
\caption{Outage rates achieved by the considered    schemes for different channel models. $M=30$, $|\mathcal{S}_1|=3$, $|\mathcal{S}_2|=300$, $N_q=2$, $r_1=40$m, $r_y=r_1$, $\alpha=3$, $R_0=1$ bit per channel use (BPCU), and $R_1=1.5$ BPCU.  }
        \label{Fig1} \vspace{-1.5em}
\end{figure}

\begin{figure}[t]
\begin{center}
\includegraphics[width=0.35\textwidth]{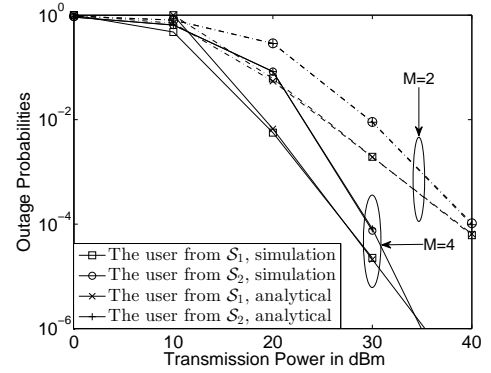}\vspace{-1em}
\end{center}
\caption{Outage probabilities achieved by the considered   schemes for the Rayleigh fading channel model. $|\mathcal{S}_1|=1$, $|\mathcal{S}_2|=M$, $N_q=2$,  $r_1=40$m, $r_y=\frac{r_1}{2}$, $\alpha=3$, $R_0=1$ BPCU, and $R_1=1$ BPCU.  }\vspace{-1.5em}
        \label{Fig2}
\end{figure}

\vspace{-1em}
\section{Numerical Results}
In this section, the performance of the proposed NOMA-MIMO scheme is   evaluated by using computer simulations. The NOMA power coefficients are set as $\alpha_{0,k}^2=3/4$ and $\alpha_{1,k}^2=1/4$. The noise power is $-30$dBm.  Without loss of generality, assume that the users in $\mathcal{S}_1$ lie  on a circle with radius $r_y$, where the base station is located at its center.

Note that for the NOMA scheme, two users are served on each beam, and they have their own   target data rates, $R_0$ and $R_1$, respectively. However, for the scheme without NOMA, a single user is served on each beam. For a fair comparison, the user's targeted data rate is $R_0+R_1$ for the case without NOMA. In Fig. \ref{Fig1},  two types of channel models, namely Rayleigh fading and the mmWave model, are considered. As can be observed from the figure, the use of NOMA can result in a significant performance gain compared to the scheme without NOMA. For example, when the transmission power of the base station is $15$dBm, the use of NOMA can offer a rate improvement of $3$ bits per channel use (BPCU) over the conventional MIMO scheme, for both considered  channel models. In Fig. \ref{Fig2}, the diversity gain achieved by the  NOMA scheme is studied. To facilitate this diversity analysis, we set $|\mathcal{S}_2|=M$, which means that the diversity gains for the users in $\mathcal{S}_1$ and $\mathcal{S}_2$ are $\frac{M+1}{2}$ and $M$, respectively. If perfect analog beamforming is used, a diversity gain of $M$ is achievable for the user in $\mathcal{S}_1$ due to FRAB. From the figure, one can clearly observe the loss of  diversity gain for the user from $\mathcal{S}_1$. Note that  the curves for the analytical results match those of the simulation results, which verifies the accuracy of our analysis.

\section{Conclusions}
In this letter, NOMA has been proposed  as a means to mitigate the reduced degrees of freedom induced by    FRAB in massive MIMO and mmWave networks. The developed analytical and simulation results have demonstrated the superior performance of the proposed NOMA scheme.

\appendices
\section{Proof of Proposition \ref{proposition}}
Recall that the probability density function (pdf) of a folded normally distributed variable is $
 f_{z_m}(x) =  \frac{2}{\sqrt{\pi}}e^{-x^2}$.
 Therefore the CDF of the square of the sum of $z_m$  is given by
 \begin{align}
 F_{z_{\sum},M}(z)
 =& \frac{2^M}{\pi^{\frac{M}{2}}}\underset{\sum_{m=1}^M x_m<\sqrt{z}}{\int \cdots \int}  \prod_{m=1}^{M} e^{-x_m^2}dx_m,
 \end{align}
 where the subscript $M$ is added to show that the CDF is a function of $M$, and to facilitate the following analysis.

 The proposition can be proved by using the inductive method. For the case  $M=1$, the approximate  expression in the proposition can be simplified as follows: \begin{align}\label{csd}
 F_{z_{\sum,1}}(z)
\approx& \frac{2}{\pi^{\frac{1}{2}}}z^\frac{1}{2}.
 \end{align}
  By calculating the integral of the pdf of $z_m$ and applying $e^{-x}\approx 1-x$ for $x\rightarrow 0$, one can verify that \eqref{csd}  is a valid approximate  expression for the CDF.

 Assuming  that the approximation is correct for the case of $M=n$,  $F_{z_{\sum, n+1}}(z)$ can be expressed  as follows:
 \begin{align}
 F_{z_{\sum, n+1}}(z)
 =& \frac{2^{n+1}}{\pi^{\frac{n+1}{2}}}\underset{\sum_{m=1}^{n+1} x_m<\sqrt{z}}{\int \cdots \int}  \prod_{m=1}^{n+1} e^{-x_m^2}dx_m
 \\ \nonumber
 =& \frac{2^{n+1}}{\pi^{\frac{n+1}{2}}}\int^{\sqrt{z}}_{0}\underset{\sum_{m=1}^{n} x_m<\sqrt{z}-x_{n+1}}{\int \cdots \int}  \prod_{m=1}^{n} e^{-x_m^2}dx_m \\ \nonumber & \times e^{-x_{n+1}^2}dx_{n+1}.
 \end{align}
 By using $ F_{z_{\sum,n}}(z)
\approx \frac{2^n}{\pi^{\frac{n}{2}}}\frac{(\sqrt{z})^n}{n!}$,  $F_{z_{\sum, n+1}}(z)$ can be approximated   as follows:
  \begin{align}
 F_{z_{\sum, n+1}}(z)\approx & \frac{2^{n+1}}{\pi^{\frac{n+1}{2}}}\int^{\sqrt{z}}_{0}\frac{(\sqrt{z}-x_{n+1})^n}{n!}  e^{-x_{n+1}^2}dx_{n+1}\\ \nonumber
 \approx & \frac{2^{n+1}}{\pi^{\frac{n+1}{2}}}\int^{\sqrt{z}}_{0}\frac{(\sqrt{z}-x_{n+1})^n}{n!}  dx_{n+1}
 \\ \nonumber
 = & \frac{2^{n+1}}{\pi^{\frac{n+1}{2}}} \frac{(\sqrt{z })^{n+1}}{(n+1)!}.
 \end{align}
 Therefore, the approximate expression is correct for   $M=n+1$, and the proof is complete via induction.

 \section{Proof for Lemma \ref{lemma1}}
 With FRAB, the user's effective channel gain can be expressed as follows:
\begin{align}
|\mathbf{h}_k^H\mathbf{f}_k|^2
&= \left|\sum^{M}_{m=1} h_{k,m}  e^{-j\frac{(i^*_{k,m}-1)2\pi}{N_q}}  \right|^2.
\end{align}

When $N_q=2$,  we  separate the real and imaginary parts of the user's channel as follows:
\begin{align}
|\mathbf{h}_k^H\mathbf{f}_k|^2 &= \left|\sum^{M}_{m=1} (h_{k,m,real}+jh_{k,m,imag}) \bar{f}_{i^*_{k,m}}\right|^2\\ \nonumber
&= \left|\sum^{M}_{m=1} (|h_{k,m,real}|+j\text{sign}(h_{k,m,real})h_{k,m,imag})  \right|^2,
\end{align}
where $\text{sign}(\cdot)$ is the sign operation.
This  separation leads to the following expression:
\begin{align}
|\mathbf{h}_k^H\mathbf{f}_k|^2
=& \left|\sum^{M}_{m=1} \bar{h}_m \right|^2  +\left|\bar{h}_0 \right|^2,
\end{align}
where $\bar{h}_0=\sum^{M}_{m=1} \text{sign}(h_{k,m,real})h_{k,m,imag} $ and $\bar{h}_m=|h_{k,m,real}|$.
Note that $ \text{sign}(h_{k,m,real})$ is independent of $|h_{k,m,real}| $, because the phase and the amplitude of a complex Gaussian random  variable are independent. Therefore, $\left|\sum^{M}_{m=1} \bar{h}_m \right|^2 $ and $\left|\bar{h}_0 \right|^2$ are independent.

Define $z_0=|\bar{h}_0|^2$ whose  pdf  can be easily obtained as follows. Recall that the sum of i.i.d. Gaussian random variables is still a Gaussian variable, which means $\bar{h}_0$ is  Gaussian with  zero mean and variance $\frac{M}{2(1+d_{yk}^\alpha)}$. Therefore,  the CDF of $z_0$ is given by
\begin{align}\label{cdfx1}
F_{z_0}(z) = \frac{\gamma\left(\frac{1}{2}, \frac{(1+d_{yk}^\alpha)z}{M}\right)}{\Gamma\left(\frac{1}{2}\right)},
\end{align}
and the pdf is $f_{z_0}(z)=\frac{dF_{z_0}(z)}{dz}$, where $\gamma(x,y)$ denotes the incomplete Gamma function.

Therefore, the CDF of $|\mathbf{h}_k^H\mathbf{f}_k|^2 $ can be expressed  as follows:
\begin{align}
F_{|\mathbf{h}_k^H\mathbf{f}_k|^2 }(y) &=\underset{x+z<y}{\int\int}f_{z_0}(z)f_{\bar{z}_{\sum}}(x)dzdx.
\end{align}

According to Proposition \ref{proposition}, when $z\rightarrow 0$,  the    pdf of $\bar{z}_{\sum}\triangleq \left|\sum^{M}_{m=1} \bar{h}_m \right|^2 $ can be approximated  as follows:
\begin{align}
f_{\bar{z}_{\sum}}(x) \approx \frac{2^{M-1} (1+d_{yk}^\alpha)^{\frac{M}{2}}}{\pi^{\frac{M}{2}}}\frac{x^\frac{M-2}{2}}{(M-1)!}.
\end{align}
  Therefore, when $y\rightarrow 0$, we have the following approximation:
\begin{align}
F_{|\mathbf{h}_k^H\mathbf{f}_k|^2 }(y)  =& \int^{ {y}}_{0}f_{\bar{z}_{\sum}}(x)\int^{y-x}_{0}f_{z_0}(z)dzdx\\ \nonumber
\approx& \int^{ {y}}_{0}\frac{2^{M-1}(1+d_{yk}^\alpha)^{\frac{M}{2}}}{\pi^{\frac{M}{2}}}\frac{x^\frac{M-2}{2}}{(M-1)!}F_{z_0}(y-x) dx.
\end{align}
Note that when $x\rightarrow 0$, the CDF in \eqref{cdfx1} can be approximated as follows:
\begin{align}
F_{z_0}(z) \approx  \frac{2(1+d_{yk}^\alpha)^{\frac{1}{2}}z^{\frac{1}{2}}}{M^{\frac{1}{2}}\Gamma\left(\frac{1}{2}\right)}.
\end{align}
Therefore, the following approximation can be obtained:
\begin{align}\nonumber
F_{|\mathbf{h}_k^H\mathbf{f}_k|^2 }(y)
\approx&  \frac{2^{M}(1+d_{yk}^\alpha)^{\frac{M+1}{2}} \int^{ {y}}_{0}x^\frac{M-2}{2}(y-x)^{\frac{1}{2}} dx}{\pi^{\frac{M}{2}}(M-1)!M^{\frac{1}{2}}\Gamma\left(\frac{1}{2}\right)} .
\end{align}
Using the definition of  the Beta function in the above expression, the lemma is proved.

 \bibliographystyle{IEEEtran}
\bibliography{IEEEfull,trasfer}

  \end{document}